\documentclass[12pt,a4paper,twoside]{article}
\usepackage{vlad_iaea2012_mod,color}
\usepackage{graphicx}
\usepackage{wrapfig}
\usepackage{amssymb}

\paper{IAEA-CN-286/TH/P5-820}{I. CHAVDAROVSKI et al.\, IAEA-CN-286/TH/P5-820}	
\begin{document}

\title{LINEAR KINETIC EFFECTS OF CORE PLASMA ON \\
LOW FREQUENCY ALFV\'EN AND ACOUSTIC  \\
EIGENMODES IN TOKAMAKS
}\\

\author{I. CHAVDAROVSKI\\
Korea Institute of Fusion Energy,
Daejeon, Republic of Korea \\
{Email: chavdarovski@gmail.com}}

\author{F. ZONCA\\
C.R. ENEA Frascati, Rome, Italy}

\author{L. CHEN\\
Dept. of Physics and Astronomy, Univ. of California,
Irvine CA, U.S.A.\\
Institute for Fusion Theory and Simulation, Zhejiang University,
Hangzhou, PR China
}

\linespread{1} \small \normalsize
\begin{abstract}

\hspace{1cm} The resonant and non-resonant effects of core plasma on the excitation of
low frequency modes with $\omega < \omega_{BAE}$, such as
Beta-induced Alfv\'en Acoustic Eigenmodes (BAAEs) and Kinetic Ballooning Modes (KBMs) are examined in the
framework of the generalized fishbone-like dispersion relation. The
formalism of the fishbone-like equation contains all the necessary ingredients to describe
the features of these low frequency fluctuations, and explain experimental findings.
Core plasma properties (diamagnetic frequency and precession resonance with trapped ions)
strongly affect the excitation of the modes, and in the case of BAAEs
more effectively than the energetic particles. The diamagnetic frequency of the core plasma also contributes to
the coupling of the BAAEs with the KBMs, thus affecting the excitation and polarization of both modes.
Energetic particles can still provide a non-resonant drive to some of the low frequency modes.
\end{abstract}

\linespread{1.} \small \normalsize

\section{INTRODUCTION}
\label{sec:intro}

Electromagnetic modes with frequencies of the order of thermal ions periodic motion $\omega_{Ti}=(T_i/m_i)^{1/2}/(q R_0)$ and $\omega_{Bi} \sim \epsilon^{1/2} \omega_{Ti}$ and precessional motion of trapped particles $\overline \omega_{Di}$, as well as their diamagnetic frequency $\omega_{*pi}=(T_ic/e_sB)(\textbf{k} \times \textbf{b}) \cdot \nabla (T_i) /n_i\,$, have been observed on several machines, such as DIII-D and HL-2A tokamaks. One of these modes named
Beta-induced Alfv\'en acoustic eigenmode (BAAE) has been theoretically described by MHD theory~\cite{gorelenkov07}
and shown to have a mixed Alfv\'enic and acoustic polarization and frequency $\omega_{BAAE} \sim (T_e/T_i)^{1/2} \omega_{Ti}$. Several experiments have reported resonant energetic particle (EP) excitation of low frequency BAAEs,
and wrongly identified them as a cause of EP losses, on a basis of their frequency range.
However, theoretical analysis shows that energetic particles are not efficient in resonantly driving the
BAAEs~\cite{chen2017}. Moreover, the BAAEs are heavily damped by thermal ions and parallel electric field~\cite{Zonetal2010,chavdar2014}, and not easy to excite. These modes are however coupled with the Kinetic Ballooning Mode (KBM) branch of low frequency Alfv\'enic oscillations~\cite{TsaiChen} with frequency close to $\omega_{*pi}\,$,
but significantly smaller damping rate. Diamagnetic effects of core plasma contribute to the coupling of the BAAEs and KBMs,
thus affecting the excitation and polarization of both modes, and in the case of BAAEs significantly reducing the damping rate~\cite{chavdar2014}. Precession resonance with trapped ions which occurs in this frequency range, is expected to have a more important role for the excitation of the low frequency modes than energetic particles.

Additionally, non-resonant effects of the background (or hot) plasma
might provide enough drive to destabilize low frequency modes, in which case
 the mode frequency is not determined by particle resonance,
but by the thermal plasma diamagnetic frequency $\omega_{*pi}$~\cite{Heidbrink21}.
The modes previously recognized as BAAEs in experiments are much more likely to be excited in this way,
and hence should be identified as reactive modes. A significant number of experiments on DIII-D tokamak~\cite{Heidbrink21}
have shown this type of modes appearing in a "Christmas lights" pattern due to the time evolution of the
$q$-profile which pases through several rational numbers $q=m/n$, with $m$ and $n$ being the poloidal and toroidal numbers, respectively. Common feature of these modes, labeled Low Frequency Modes (LFMs) is that they appear near $q_{min}$ of a
reversed shear profile, and they are driven by high electron temperature $T_e$, but modest $\beta_i$.
Their polarization is Alfv\'enic and the instability is reactive in nature.

The closeness of the modes to the ion resonant frequencies has prompted a development of a kinetic treatment~\cite{Zonetal2010}
of both circulating~\cite{ZC96} and trapped particles~\cite{chavdar2009,chavdar2014} to properly describe
these low frequency specimen. Due to the large number of activities in this frequency range
the identification of the mode has to include the frequency, damping rate and its polarization.
The latter is however difficult to determine experimentally, which is one of the reasons why analytical tools such as the
Generalized fishbone-like dispersion relation (GFLDR) is useful for systematic study of these low frequency modes.
In this framework the dispersion relation of the mode is given by \cite{TsaiChen}
\begin{equation}\label{e:fish}
i \Lambda(\omega) = \delta \bar{W}_f + \delta \bar{W}_k \;\; , \label{eq:fishlike}
\end{equation}
which is a unifying picture for various Alfv\'enic fluctuations, as well as Energetic particle continuum modes (EPMs).
Here, $\Lambda$ is the general inertia and represents the
physics inside the inertial layer, while $\delta \bar{W}_f$ and $\delta \bar{W}_k$ are the ideal region
background MHD and energetic particle kinetic contribution, respectively.
The scope of this work is to show that the GFLDR in its most general form contains all the ingredients necessary
to describe the behavior of the low frequency modes and their mutual interaction.

\section{THEORETICAL MODEL}
\label{sec:theory}

We adopt the ballooning
formalism in the space of the extended poloidal angle $\theta$, for low $\beta=O(\epsilon^{2})$
axisymmetric $(s,\alpha)$ plasma equilibrium, with $s = rq'/q$ being the finite magnetic shear and $\alpha = - R_0 q^2 \beta'$ the balloning drive. We describe the plasma with three fluctuating scalar fields: $\delta \phi$, $\delta \psi$ and $\delta B_\parallel$.
Eq.~(\ref{e:fish}) is obtained by asymptotic
matching of the "singular" inertial layer and "regular" ideal region solutions of the eigenmode (vorticity) equation for $\delta \psi$ at the modes rational surface~\cite{TsaiChen}.
In the inertial layer the fields vary
on a short scale $\theta_0 \sim O(1)$ and on a long scale $\theta_1 \sim (\beta^{1/2})$, while their
governing equations in the long wavelength limit are the vorticity equation~\cite{TsaiChen,ZC96}
\begin{eqnarray}\label{e:vort}
 B \textbf{b} \cdot \nabla \left[{1 \over B } {{k_\perp}  ^2\over {k_\vartheta} ^2} \textbf{b}\cdot \nabla \delta\psi \right] +  {\omega^2 \over v_A^2} \left(1- {\omega_{\ast pi} \over \omega }\right) \frac{k_\perp^2}{k_\vartheta^2}  \delta\phi + {\alpha \over q^2 R^2} g (\theta)\delta \psi = \Bigl \langle \frac{4\pi e}{k_\vartheta^2 c^2} \omega \omega_{di} \delta K_i \Bigr \rangle \, ,
\end{eqnarray}
and quasi-neutrality condition
\begin{equation}\label{e:quas}
\left( 1+\frac{1}{\tau}\right)(\delta\phi - \delta\psi) = \frac{T_i}{ne}\langle \delta K_i  - \delta K_e  \rangle \, .
\end{equation}
Here $\delta K_i$  and $\delta K_e$ represent the compressional parts of the perturbed particle distribution function $\delta f_s$, for $s=i,e\,$, while $\langle (...)\rangle=\int d\textbf{v}(...)$ denotes integration in velocity space. The perpendicular $k_\perp$ and poloidal $k_{\theta}$ wave numbers satisfy $k_\perp^2/k_\theta^2 = 1 + (s\theta-\alpha \sin \theta)^2$, while $g(\theta)=\cos \theta +[s\theta-\alpha \sin \theta \,] \sin \theta\,$, $\tau = T_e/T_i$ and $n_e = n_i = n$. Further, we will use the more convenient variables $\delta \Phi=(k_\perp/k_\theta)\, \delta \phi$ and
$\delta \Psi=(k_\perp/k_\theta)\, \delta \psi$, which represented in series of powers of $\beta^{1/2}$ give (see~\cite{ZC96} and~\cite{chavdar2009}) $\delta \Psi^{(0)}=\delta \Psi^{(0)}(\theta_1)$, $\delta \Psi^{(1)}=0$ and
$\delta \Phi^{(1)} \simeq \delta \Phi^{(0)}+ \delta \Phi_s (\theta_1) \sin \theta_0$. Here, $\delta \Phi_s (\theta_1) \sin \theta_0$ is the sinusoidal a.c. term of the perturbed electric potential, while $\delta \Phi_c (\theta_1) \cos \theta_0$ was dropped, since $\delta \Phi_c \ll \delta \Phi_s$~\cite{ZC96,chavdar2009}.

For trapped ions the distribution function is written as~\cite{chavdar2009}
\begin{eqnarray}\label{e:deltai}
 \delta K_i= &-&Qf_{0i} \left(\frac{e}{m} \right)_i \frac{1}{\omega-\overline\omega_{di}}\frac{k_\theta}{k_\perp} \left[ \frac{\overline\omega_{di}}{\omega} \delta \Psi^{(0)}+\delta\Phi^{(0)}-\delta\Psi^{(0)}\right] \nonumber
\\&-&Qf_{0i} \left(\frac{e}{m} \right)_i \frac{1}{(\omega-\overline\omega_{di})^2-\omega_b^2}\frac{k_\theta}{k_\perp} \left[{\overline\omega_{di}} \delta \Phi^{(0)} \xi+(\omega-\overline{\omega}_{di})\delta\Phi_{s}\right] \sin \theta \,\, .
\end{eqnarray}
The second term contains the $\overline\omega_{d}\pm \omega_{b}$ resonances and is an essential contribution to the inertial layer dynamics, while the first one containing $\overline{\omega}_{di}$ resonance is found to be of order $\xi=k_\perp/k_\theta$ smaller than the second one. The ratio $k_\perp/k_\theta$ is large in the inertial layer, but of $O(1)$ in the ideal region, hence the precession resonance will not contribute to the inertial layer dynamics, aside from the quasineutrality condition. In the ideal region, however we can expect the thermal ion precession resonance to appear on the right hand side of Eq.(\ref{e:fish}), and have effect on modes of the order of $\overline{\omega}_{di}$, i.e. low frequency modes.

The quasineutrality condition in zeroth order gives~\cite{chavdar2014} $\delta \Phi^{(0)}=I_\Phi(\omega,\omega_{Di,e},\omega_{*i,e}) \delta \Psi^{(0)}$, where $I_\Phi(\omega,\omega_{Di,e},\omega_{*i,e})$ represents a deviation from the ideal MHD limit $\delta E_\parallel=0$, obtained for $I_\Phi=1$. For most of the spectrum of the Alfv\'enic fluctuations, except frequencies near $\overline{\omega}_{D,ei}$, we have $I_\Phi\simeq 1$ thus recovering the ideal MHD condition. Around the specific precession frequencies of the ions and electrons, $I_\Phi\neq1$ which gives a finite parallel electric field as a result of the slow precessional motion of the electrons and ions in opposite directions. This electric field might have important implications on the collisionless micro-tearing effects, but also contribute to the loss of particles on the trapped/passing border.

Next order ($\beta^{1/2}$) quasineutrality condition gives $\delta \Phi_s (\theta_1)=S(\omega,\omega_{Bi},\overline{\omega_{Di}},\omega_{Ti})\xi \delta \Phi^{(0)}$, which recalling $\delta \Psi^{(1)}=0$ makes the function $|S|$ a measure of how much the polarization deviates from pure Alfv\'enic due to the parallel a.c. electric field. The polarization term $S$ was calculated from the quasineutrality condition~\cite{chavdar2009} to obtain:
\begin{equation}\label{e:S}
\hskip -2em S=-\frac{  N_1(\frac{\omega}{\omega_{Ti}}) + \Delta N_1(\frac{\omega}{\omega_{Ti}}) +\sqrt{2\epsilon}P_2(\frac{\omega}{ \overline \omega_{Di}},
\frac{\omega_{Bi}}{ \overline \omega_{Di}}) }{1+\frac{1}{\tau}+ D_1(\frac{\omega}{\omega_{Ti}}) + \Delta D_1(\frac{\omega}{\omega_{Ti}}) +
\sqrt{2\epsilon}\left[P_1(\frac{\omega}{ \overline \omega_{Di}},\frac{\omega_{Bi}}{ \overline \omega_{Di}})-P_2(\frac{\omega}{ \overline \omega_{Di}},\frac{\omega_{Bi}}{ \overline \omega_{Di}})\right]}
\,  ,
\end{equation}
where $P_1(\omega/\overline\omega_{Di},\omega_{Bi}/\overline\omega_{Di})$ and $P_2(\omega/\overline\omega_{Di},\omega_{Bi}/\overline\omega_{Di})$ come from the trapped particles dynamics:
$$P_1(\omega/\overline\omega_{Di},\omega_{Bi}/\overline\omega_{Di}) = - 2 \frac{\omega^2}{\overline\omega_{Di}^2}
\left[(1-\frac{\omega_{*n}}{\omega}+\frac{3}{2}\frac{\omega_{*T}}{\omega}) G_2 - \frac{\omega_{*T}}{\omega} G_4\right]\, ,$$
$$P_2(\omega/\overline\omega_{Di},\omega_{Bi}/\overline\omega_{Di}) = - 2  \frac{\omega}{\overline\omega_{Di}}
\left[(1-\frac{\omega_{*n}}{\omega}+\frac{3}{2}\frac{\omega_{*T}}{\omega}) G_4 - \frac{\omega_{*T}}{\omega} G_6\right]\, ,$$
and we have denoted
$$G_n=\frac{1}{\pi^{1/2}} \int_{-\infty}^\infty \frac{e^{-x^2} x^n}{(\omega/\overline\omega_{Di}- x^2)^2-(\omega_{Bi}/\overline\omega_{Di})^2 x^2}\,dx \, ,$$
for $n=2,4,6,8$. The $G_n$ integrals yield~\cite{chavdar2009}
\begin{eqnarray}
G_2 &=& \frac{\overline\omega_{Di}/\omega_{Bi}}{\Omega_1+\Omega_2} \left[ \Omega_1 Z(\Omega_1) - \Omega_2 Z(\Omega_2) \right] \;\; ,\\
G_4 &=&\frac{\overline\omega_{Di}/\omega_{Bi}}{\Omega_1+\Omega_2} \left[ \Omega_1^2 - \Omega_2^2 + \Omega_1^3 Z(\Omega_1) - \Omega_2^3 Z(\Omega_2) \right] \;\; ,\nonumber \\
 G_6 &=&\frac{\overline\omega_{Di}/\omega_{Bi}}{\Omega_1+\Omega_2} \left[ (1/2) (\Omega_1^2 - \Omega_2^2) + \Omega_1^4 - \Omega_2^4 + \Omega_1^5 Z(\Omega_1) - \Omega_2^5 Z(\Omega_2) \right] \;\; ,\nonumber \\
G_8 &=&\frac{\overline\omega_{Di}/\omega_{Bi}}{\Omega_1+\Omega_2} \left[ (3/4) (\Omega_1^2 - \Omega_2^2) + (1/2) (\Omega_1^4 - \Omega_2^4) + \Omega_1^6 - \Omega_2^6 + \Omega_1^7 Z(\Omega_1) - \Omega_2^7 Z(\Omega_2) \right], \nonumber
\label{eq:fgnd}
\end{eqnarray}
where $Z(x)=1/ \sqrt\pi \int_{-\infty}^{\infty} e^{-y^2}/(y-x)\,dy$ is the plasma dispersion function that contains the $\omega=\overline\omega_{Di}\pm \omega_{Bi}$ resonances in the parameters $\Omega_1$ and $\Omega_2$ given by
$$\Omega_1=\frac{\frac{\omega_{Bi}}{\overline\omega_{Di}}+\sqrt{(\frac{\omega_{Bi}}{\overline\omega_{Di}})^2+
4\frac{\omega}{\overline\omega_{Di}}}}{2} \,\, {\rm and}\,\,
\Omega_2=\frac{-\frac{\omega_{Bi}}{\overline\omega_{Di}}+\sqrt{(\frac{\omega_{Bi}}{\overline\omega_{Di}})^2+
4\frac{\omega}{\overline\omega_{Di}}}}{2}\,\, . $$
In Eq.~(\ref{e:S}), the functions
\begin{equation}\label{e:D}
D_1(x)=x \left(1-\frac{\omega_{*ni}}{\omega}\right)Z(x)- \frac{\omega_{*Ti}}{\omega}x [x+(x^2-1/2)Z(x)]
\end{equation}
and $N_1(x)= 2 (\overline \omega_{Di}/{\omega_{Ti}} ) N(x) \,, $
 with
\begin{equation}\label{e:N}
\hskip -2em N(x)= \left(1-\frac{\omega_{*ni}}{\omega}\right)[x+(1/2+x^2)Z(x)]- \frac{\omega_{*Ti}}{\omega}[x(1/2+x^2)+(1/4+x^4)Z(x)]\,\, ,
\end{equation}
come from the well circulating particles dynamics~\cite{ZC96}. These functions contain the $\omega=\omega_{Ti}$ resonance and are strictly valid for well circulating particles. In order to extend their validity virtually to $\omega\simeq 0$,
correction functions
\begin{equation}
\hspace*{-2cm}\Delta D_1 (x) = \frac{x}{\pi^{1/2}} \int_0^\infty e^{-y} \ln  \left(\frac{x+\sqrt{2\epsilon y}}{x-\sqrt{2\epsilon y}}\right)  \left[ 1-\frac{\omega_{*ni}}{\omega} - \frac{\omega_{*Ti}}{\omega} \left( y - \frac{3}{2} \right)  \right] dy \label{e:dd1}
\end{equation}
and
\begin{equation}
\Delta N_1 (x) = \frac{\overline\omega_{Di}/\omega_{Ti}}{\pi^{1/2}} \int_0^\infty y e^{-y}  \ln \left(\frac{x+\sqrt{2\epsilon y}}{x-\sqrt{2\epsilon y}}\right) \left[ 1-\frac{\omega_{*ni}}{\omega} - \frac{\omega_{*Ti}}{\omega} \left( y - \frac{3}{2} \right) \right] dy \;\; , \label{e:dn1}
\end{equation}
were added~\cite{chavdar2009}. The corrections also account for the finite trapped particle fraction and reduce the circulating population by $\sqrt{2\epsilon}$.
This model is a mixture of deeply trapped and well circulating particles,
so some error is always expected when working in the low frequency regime. However, it has been shown that this reduced model
recovers well the low and high frequency behavior of hot plasmas~\cite{chavdar2009}, and even gives new insights
which are consistent with the experiments.

The vorticity equation expanded to the second order in the inertial layer
reduces to the form $(\partial^2 /\partial \theta_1^2) \delta \Psi^{(0)} + \Lambda^2 \delta \Psi^{(0)}=0$, where the general expression of $\Lambda^2$ can be written as~\cite{chavdar2009}
\begin{equation}\label{e:final}
\Lambda^2/I_\Phi = \frac{\omega^2}{\omega_A^2} \left ( 1- {\omega_{\ast pi} \over \omega} \right)+ \Lambda^2_{cir} +\Lambda^2_{tra} \; ,
\end{equation}
with $\Lambda^2_{cir}$\cite{ZC96} and $\Lambda^2_{tra}$\cite{chavdar2009} being the circulating and trapped particle contributions, respectively. Here
\begin{equation}\label{e:Ltra}
 \Lambda^2_{tra} = \frac{\omega^2\omega_{Bi}^2}{\omega_A^2 \overline \omega_{Di}^2}\frac{q^2}{\sqrt{2\epsilon}}
\left[P_3+ (P_2-P_3)S(\omega,\overline \omega_{Di},\omega_{Bi},\omega_{Ti}) \right]
\end{equation}
and $$P_3 = - 2\left[(1-\frac{\omega_{*n}}{\omega}+\frac{3}{2}\frac{\omega_{*T}}{\omega})G_6-\frac{\omega_{*T}}{\omega}G_8 \right] \,\, .$$
The circulating particle term in Eq.~(\ref{e:final}) contains the well circulating response~\cite{ZC96} and the previously mentioned corrections
\begin{eqnarray*}& & \hspace*{-1.5cm} \Lambda^2_{cir} =
q^2\frac{\omega \omega_{Ti}}{\omega_A^2} \left[\left(1-\frac{\omega_{\ast ni}}{\omega}\right) \left( F\left(\frac{\omega}{\omega_{Ti}}\right) + \Delta F\left(\frac{\omega}{\omega_{Ti}}\right) \right) - \frac{\omega_{\ast Ti}}{\omega}\left( G\left(\frac{\omega}{\omega_{Ti}}\right)  \right. \right. \\ & & \left. \left. \hspace*{-2em} + \Delta G\left(\frac{\omega}{\omega_{Ti}}\right) \right) + \frac{\omega\omega_{Ti}}{4\overline\omega_{Di}^2}\left( N_1\left(\frac{\omega}{\omega_{Ti}}\right) + \Delta N_1\left(\frac{\omega}{\omega_{Ti}}\right) \right)  S(\omega,\overline \omega_{Di},\omega_{Bi},\omega_{Ti}) \right]
\; , \end{eqnarray*}

which, in the $\epsilon \rightarrow 0$ limit reduces to the well circulating particle equations of Ref.~\cite{ZC96} with:
\begin{eqnarray}
\hskip -4em F(x) & =& x \left( x^2 + 3/2 \right) +  \left( x^4 + x^2 + 1/2
\right) Z(x) \, ,\nonumber \\
\hskip -4em G(x) & =& x \left( x^4 + x^2 + 2 \right) +  \left( x^6 + x^4/2
+ x^2 + 3/4 \right) Z(x) \, . \label{eq:fgnd}
\end{eqnarray}
The corrections are given by~\cite{chavdar2009}:
\begin{equation}
\Delta F (x) = \frac{1}{\pi^{1/2}} \int_0^\infty e^{-y} \ln  \left(\frac{x+\sqrt{2\epsilon y}}{x-\sqrt{2\epsilon y}}\right)  \frac{y^2}{4} dy \;\; , \label{e:df}
\end{equation}
and
\begin{equation}
\Delta G (x) = \frac{1}{\pi^{1/2}} \int_0^\infty e^{-y} \ln  \left(\frac{x+\sqrt{2\epsilon y}}{x-\sqrt{2\epsilon y}}\right)  \frac{y^2}{4} \left( y - \frac{3}{2} \right)  dy. \label{e:dg}
\end{equation}

The fluid limit ($\omega \gg \omega_{ti}$ and $\omega_* \rightarrow 0$) of Eq.~(\ref{e:final}) is
$ \Lambda^2= [\omega^2-\omega_{BAE}^2(1+\delta(\tau,q,\omega)]/\omega_A^2\,, $
which contains the known BAE accumulation point $\omega^2_{BAE}=2 T_i(7/4+\tau)/(m_i R_0^2) $,
with a higher order correction $\delta(\tau,q,\omega)$~\cite{RMP}. The Beta-induced Alfv\'en eigenmode is located in
the gap below the continuum and its frequency is a function of $\omega_{*pi}$~\cite{ZC96,chavdar2014}.
At low frequency the same equation gives the Rosenbluth-Hinton inertia enhancement~\cite{Rosenbluth,ZoncaIAEA}
\begin{equation}\label{e:RR}
\Lambda^2= \frac{\omega^2}{\omega_A^2} \left ( 1- {\omega_{\ast pi} \over \omega}\right)(1+1.6 q^2 \epsilon^{-1/2}+ 0.5 q^2) \,\,,
\end{equation}
which contains the known ion polarization term, in addition to the ion compression in toroidal geometry,
i.e. neoclassical enhancement. The term $I_\phi$ in Eq.~(\ref{e:final}) acts as an additional inertia enhancement due to the
slow precessional motion of ions and electrons around the torus in opposite directions.

Several different derivations of $\delta \bar{W}_f$ are known, one of which shown here~\cite{riri2014}
\begin{equation}\label{e:dWf}
\delta \bar{W_f} \simeq \frac{\pi}{|s|} \left[\frac{s^2}{4}-\frac{3 \alpha |s|}{2}+\frac{5 \alpha^2 s^2}{32}+\frac{45 \alpha^4}{128}-(1+\frac{\alpha}{2}) e^{-\frac{1}{|s|}}\right] \; ,
\end{equation}
is valid for $|s|\sim \alpha^2$ and small $\alpha=(1+\tau)(1+\eta_i) q^2 \beta_i/\epsilon_{ni}+q^2 \beta_E/\epsilon_{nE}$, where $\epsilon_{ns}=L_{ns}/R_0$ for $s=i,E$, and $L_{ns}$ are the density gradient scales of the thermal and energetic ions. The $\delta \bar{W}_f$ term accounts for the thermal electron temperature in $\tau$ and non-resonant EP effects ($\beta_E$), as well as the sharp spacial gradient of EPs through $L_{nE}$.
Detailed discussion about the $\delta \bar{W}_k$ term can be found in Ref.~\cite{zonca2014}. Here, due to the low
frequency we focus on the resonant interaction of the modes with the particle precessional motion, neglecting the bounce frequency resonance.
A fairly exhaustive analysis of the modes can be conducted by using the EP $\delta \bar{W}_{k}$ term of Ref.~\cite{zonca2014} for trapped ions
\begin{equation}\label{e:dWk}
\hskip -2em \delta \bar{W}_{kt}=\frac{2\pi^2}{|s|}\frac{e^2}{mc^2}q R_0 B_0 \int d\varepsilon \int
d\mu \left(\frac{\overline{\omega}_D}{k_\vartheta}\right)^2\left(1-\frac{\Delta_{b0}}{(1+\Delta_{b0}^2)^{1/2}}\right)
\frac{\tau_b \, QF_0 }{\overline{\omega}_D-\omega} \, ,
\end{equation}
where $Qf_{0}=(\omega\partial_\varepsilon+\hat{w}_\ast )_s f_{0}$, $\tau_b=2\pi/\omega_b$, while $\Delta_{b0}^2=
k_{\theta}^2 (\rho_L^2+\rho_b^2)/2$ accounts for both finite Larmor radius ($\rho_L$) and finite banana
orbit width ($\rho_b$). This term was originally derived for energetic particles~\cite{TsaiChen}, but it also applies to thermal ions and electrons
in the small FLR/FOW limit ($\rho_L,\rho_b \rightarrow 0$).
While for energetic particles the spacial gradient
term $\hat{w}_\ast$ is dominant, for thermal particles the energy derivative is of the same order.
Here, we remind the reader that for thermal electrons the frequencies in $\delta \bar{W}_{kt}$ should be replaced with $\omega_{*i}\rightarrow -\tau \omega_{*i}\,$, $\overline{\omega}_{Di}\rightarrow -\tau \overline{\omega}_{Di}$ and $\omega_{bi}\rightarrow \sqrt{\tau} \omega_{bi}$, giving another dependence of the modes on the electron temperature.
Since $\omega_{*p}$ and $\overline{\omega}_D$ for electrons and ions are of the same order, both $\delta \bar{W}_{kt}$
terms should be kept in the analysis on the righthand side as $\delta \bar{W}_{ki}+\delta \bar{W}_{ke}$.
However, due to the opposite signs of the resonant frequencies they would likely affect different modes, at different times. Eq.~(\ref{e:dWk}) for thermal particles has strong effects in low frequency regimes ($\omega \sim \omega_{*p},\overline{\omega}_D$), but not around $\omega_{BAE}$.

\section{DISPERSION RELATION OF THE LOW FREQUENCY MODES}

\begin{figure}[h!]
\begin{center}
\includegraphics[width=0.45\textwidth]{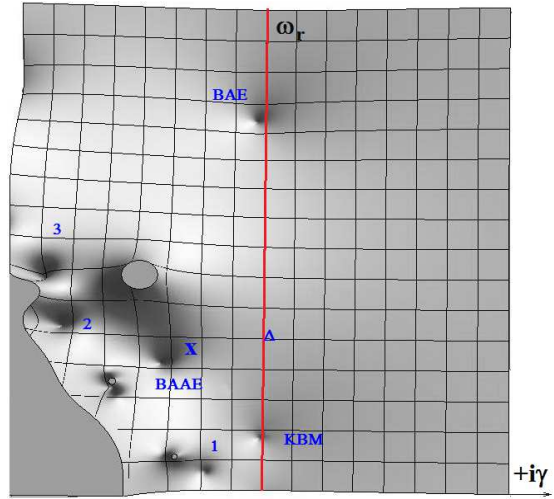}
\caption{\label{fig:fig1} Accumulation points of BAE, KBM and BAAE in complex plane for $\eta_i=0$, $\tau=1$ and $q=4/3$.
The vertical scale is $\omega_r$, horizontal $i\gamma$, thick red line $\gamma=0$. '$\Delta$' marks the BAAE for $\eta_i=3.5$ and $\Omega_*=1$, while 'x' marks the KBM for $\eta_i=1$ and $\Omega_*=0.8$.}\label{fig:1}
\end{center}
\end{figure}

The GFLDR contains all the elements of the kinetic thermal ion gap necessary to identify and study low frequency modes and their polarization, including the background plasma MHD potential, resonant and non-resonant energetic particle drive,
as well as thermal ion and electron dynamics.
Hence, the equation as roots has both the reactive low frequency instability~\cite{Heidbrink21}, as well as
several different modes in this frequency range like BAEs, EPMs, KBMs, BAAEs and MHD
modes~\cite{chen1994}.
For modes located at the vanishing shear, i.e. $q_{min}$ of a reversed shear profile,
the LHS of Eq.(\ref{eq:fishlike}) should be expanded around $s=0$, as done in Ref.\cite{zonca2014}.
The equation $\Lambda=0$ was solved in Ref.~\cite{chavdar2014} and the accumulation points were obtained for
the three modes shown in FIG.~\ref{fig:1}, without the presence of EPs.
Since the functions in $\Lambda$ are transcendental, additional roots in the negative complex plane appear, in which
case for each branch only the most unstable mode is physically relevant.
The roots numbered $1-3$ are mixed Alfv\'en-acoustic, and they do not show excitation for the parameters
considered in this work. For low $\eta_i$ and $\Omega_*$ the KBMs is marginally stable and has a
frequency close to $\omega_{*pi}$, i.e, proportional to the $n$- mode number up to values of $\omega/\omega_{ti} \sim 1.5$.
For increasing $\omega_{*pi}$ this behavior is modified due to the coupling
with the BAAE~\cite{chavdar2014} and BAE~\cite{ZC96}, and the KBM is becoming more damped.
In FIG.~\ref{fig:1} the KBM is shown for $\eta_i=1$ and $\Omega_*=0.8$ marked with an 'x' in the negative complex plane.
The coupling to the KBM is of great importance for the excitation of BAAEs since they are shown to have a large damping rate~\cite{Zonetal2010}, that only reduces for high $\eta_i$ values~\cite{chavdar2014}.

The BAAEs are modes with mixed Alfv\'en-acoustic polarization. In our model, the polarization of the modes is determined by the $S$ function of Eq.(\ref{e:S}). Modes with Alfv\'enic polarization have $\delta \Phi^{(0)}=\delta \Psi^{(0)}$ and $|S| \sim \beta^{1/2}$ or lower, while modes with significant acoustic component, such as BAAEs are identified by having $|S|\gg \beta^{1/2}$. In the fluid limit of the theory, around the BAE accumulation point $|S|\sim \beta^{1/2}$, which is also the case in the opposite limit $\omega \ll \omega_{bi}$. This explains the vanishing of the $\tau$ term in the Rosenbluth-Hinton Eq.(\ref{e:RR}). At low $\eta_i$ the KBM is purely Alfv\'enic, but for increasing $\eta_i$ and $\Omega_*$ it becomes of mixed polarization, which creates an issue with the identification of the mode as a BAAE or KBM. Detailed discussion of this problem can be found in Ref.\cite{chavdar2014}. The increase of the diamagnetic frequency can also change the polarization of the BAAE, while decreasing its damping rate. In FIG.\ref{fig:1} the BAAE mode for $\eta_i=3.5$ and $\Omega_{*}=1$ is marked with '$\Delta$'. Here, the BAAE is already unstable, but  its frequency is below the EP resonant frequencies and $\omega_{*pi}$. The mode could be driven by a non-resonant EP drive coming from $\delta \bar{W}_f+\delta \bar{W}_k$, as shown by the numerical simulations with GTC code, but this might result with a change of the polarization to Alfv\'enic~\cite{zhang2016}.
This excitation is non-resonant, i.e. BAAEs would weakly affect EP transport. Additionally, the BAAEs may be excited by the precessional motion of thermal particles through $\delta \bar{W}_k$ in Eq.(\ref{e:dWk}), confirmed by GTC simulations that show power exchange between one of these low frequency modes with the thermal ions at the precession frequency.

When discussing the accumulation points ($\Lambda$=0), the electron temperature only appears through $\tau=T_e/T_i$ in the polarization term, so when $S$ is negligible the mode is not affected by $T_e$. The KBM frequency $\omega \simeq \omega_{*pi}$ at low $\eta_i$ is not correlated with the electron temperature. Similarly the accumulation point of the BAAE at $\eta_i=0$ is not strongly affected by $\tau$, even though the usual assessment is $\omega_{BAAE} \propto (T_e/T_i)^{1/2}$. BAE frequency is strongly dependent of $\tau$ since the functions in the denominator of Eq.(\ref{e:S}) add up to $-1$, which brings forth the $1/\tau$ term. This is not the case for BAAEs, so any significant effects of $T_e$ on this branch is not coming from the accumulation point.

It is important to mention that both the terms $\delta \bar{W}_f$ and $I_{\phi}$ also contain $\tau$,
so the origin of the effects of $T_e$ and $\nabla T_e$ on the modes should be investigated carefully. In the case
of the LFM in Ref.\cite{Heidbrink21}, increasing $T_e$ drives the mode, but increasing $T_i$ and $\nabla T_i$ has stabilizing effects, while the mode frequency is closely following the $n$ mode number. The later is also valid for KBMs at low $\omega_{*pi}$. The effects of electron temperature on these low frequency modes can be fully explained when the RHS of Eq.~(\ref{e:fish}) is taken into account, in which case the mode is localized slightly off the rational surface.
The location of the accumulation point of the mode (rational surface) is determined by the mode numbers and $q$-profile, while the mode frequency itself by matching the real part of $i \Lambda$ with $Re\,(\,\delta \bar{W}_f+\delta \bar{W}_k)$. The imaginary part of $\delta \bar{W}_k$ compensates for the damping on the LHS of GFLDR, i.e. $Im (i\Lambda)=Im (\,\delta \bar{W}_k)$.

\section{CONCLUSIONS}
\label{conclusions}

The thermal particles play an important role in the dynamics of the low frequency modes located in the kinetic thermal
ion gap, even more than energetic particles themselves.
The low frequency Alfv\'enic mode closely connected with the ion diamagnetic frequency known as KBM, is easier to excite than the mixed polarization BAAE. The latter is however coupled with the KBM through the ion diamagnetic frequency, and becomes marginally unstable only for high $\Omega_*$ and $\eta_i$, in which case it could be non-resonantly excited by energetic particles. This however can change its polarization to Alfv\'enic. The precession resonance with thermal ions and electrons, as well as the effects from $\delta \bar{W}_f$ due to energetic particle $\beta$ and $\omega_*$ are worthwhile examining as a drive for some of these modes. Out of the modes mentioned in this work, only the BAEs strongly interact with EPs in
tokamak plasma, and can be deleterious to their confinement.
The LFMs are known to be driven by high electron temperature, which can be traced to the thermal part of the RHS terms in Eq.(\ref{eq:fishlike}). The GFLDR is the proper description to explain the linear excitation of the low frequency branches regardless weather they are reactive of resonant, since it contains all the relevant physics of fluctuations in the thermal ion kinetic gap. It is a general and comprehensive tool for understanding mode structures at low frequencies and explaining experimental observations.

ACKNOWLEDGMENT

This work is supported by R\&D Program through the Korean Institute of Fusion Energy (KFE)
funded by the Ministry of Science and ICT of the Republic of Korea (KFE-EN2141-7).

\end{document}